\documentclass[doublecol, nobibnotes]{epl2}
\usepackage{graphicx}
\usepackage{epsfig}
\usepackage{amsmath, amsthm, amssymb}

\newcommand{\bb}{{\bf B}}

\newcommand{\bx}{{\bf x}}
\newcommand{\vel}{{c}}
\newcommand{\bvel}{{\bf c}}

\newcommand{\bk}{{\bf k}}

\newcommand{\bdist}{{\bf g}}
\newcommand{\dist}{{g}}

\newcommand{ \dotp }{ {\mbox{\boldmath$\cdot$}} }
\newcommand{ \bnabla }{ {\mbox{\boldmath$\nabla$}} }

\newcommand{ \lam }{ \Lambda }

\title{Lattice Boltzmann Method for Electromagnetic Wave Propagation}
\author{S. M. Hanasoge\inst{1,2} \and S. Succi\inst{3,4} \and S. A. Orszag\inst{5}}
\shortauthor{S. M. Hanasoge \etal}
\institute{
\inst{1} Department of Geosciences, Princeton University, New Jersey 08544, USA \\
\inst{2} Max-Planck-Institut f\"{u}r Sonnensystemforschung, Max-Planck Stra$\beta$e 2, 37191 Katlenburg-Lindau, Germany\\
\inst{3} Istituto Applicazioni Calcolo, CNR Roma, via dei Taurini 9, 00185, Roma, Italy \\
\inst{4} Freiburg Institute for Advanced Studies, Freiburg, Germany \\
\inst{5} Department of Mathematics, Yale University, New Haven, Connecticut 06520-8283, USA
}

\abstract{
We present a new Lattice Boltzmann (LB) formulation to solve the Maxwell 
equations for electromagnetic (EM) waves propagating in a heterogeneous medium. 
By using a pseudo-vector discrete Boltzmann distribution, the scheme is shown 
to reproduce the continuum Maxwell equations. 
The technique compares well with a pseudo-spectral method at solving for two-dimensional wave 
propagation in a heterogeneous medium, which by design contains substantial contrasts 
in the refractive index. The extension to three dimensions follows naturally and, 
owing to the recognized efficiency of LB schemes for parallel computation 
in irregular geometries, it gives a powerful method to numerically simulate
a wide range of problems involving EM wave propagation in complex media.
}
\begin{document}
\maketitle

\section{Introduction}

Over the last two decades, the LB method in the Bhatnagar-Gross-Krook (BGK) approximation
(see \cite{BGK}) has
met with significant success in simulating a  broad
spectrum of complex flow phenomena \cite{BSV,AIDUN}, ranging from low-Reynolds
flows in porous media to fully developed turbulent flows in complex geometries,
multiphase flows and, more recently, relativistic flows \cite{RLB}. 
The application of LB techniques to EM wave propagation phenomena 
is comparatively far less developed.
Given the paramount role of EM phenomena in 
science and technology, including fast-emerging applications such 
as metamaterials and transformation
optics \cite{META}, it is of great interest to investigate whether LB techniques
are able to improve simulations of complex EM phenomena as they have for fluid flows. 

LB schemes for wave propagation have been proposed before in
literature, \cite{MORA,CHOPA,GUANG}, but it is only recently that such schemes
have been specifically tailored to EM phenomena.
To this end, a basic issue had to be addressed, namely the
fact that, unlike hydrodynamics, EM interactions are governed by 
an {\it anti}-symmetric field tensor, a structure that does not naturally emerge 
from standard kinetic theory.
To circumvent this problem, Mendoza et al. \cite{MENDOZA1,MENDOZA2} proposed 
a scheme wherein the electric and magnetic fields are represented by
separate discrete Boltzmann distributions, each moving along distinct
lattice directions.
Although an important conceptual advance, the resulting scheme appears
computationally intensive.

A more straightforward formulation for EM wave propagation in plasmas 
has been recently proposed by Dellar (\cite{DELLAR1}, \cite{DELLAR2}).
By promoting the discrete Boltzmann distribution from a scalar to a vector
quantity (also see \cite{CHOPA}, \cite{MARCONI}), and expressing the curl of electric field in divergence
form, Dellar developed an elegant and compact scheme in which
magnetic field emerges as the \lq \lq vector density" associated
with the vector-valued discrete Boltzmann distribution.
The Maxwell equations emerge when a Hermite-projection procedure
is applied to the resulting kinetic equation. 
However, antisymmetry of the relevant EM tensor is not 
preserved in time, so it must be enforced at each time-step. 

Motivated by Dellar's core idea of the discrete Boltzmann distribution not having to be a scalar,
we formulate a new scheme that is computationally inexpensive, inherently 
maintaining antisymmetry of the EM tensor.
 To this end, we introduce a {\it tensorial} distribution 
function, with built-in antisymmetry $g_{\alpha\beta}=-g_{\beta\alpha}$, where 
greek subscripts $\alpha,\beta$ run over spatial dimensions.
Tensor $g_{\alpha\beta}$ is in fact a {\it pseudo-vector}, entailing
only $D$ independent components in $D$-dimensional space.
Furthermore, and most importantly for practical applications,
we develop a new formulation wherein heterogeneity is 
realized through space-time-dependent permittivities and light speed, and embedded
within a source term in the Maxwell equations.
The scheme developed here permits us to address the important problem of 
EM wave propagation in strongly heterogeneous materials.
The scheme is validated by direct comparison with a
pseudo-spectral method for the case of wave propagation in both 
homogeneous and heterogeneous media.

\section{The Maxwell Equations}

We start with the Maxwell equations defined  in arbitrarily heterogeneous, charge-free media. 
In principle, we allow for frequency-dependent dielectric permittivity, which
introduces a temporal convolution, resulting in integro-differential Maxwell equations. 
For the sake of simplicity, we consider magnetic permeability as only a function of space (and therefore constant in time);
this assumption may be relaxed and frequency-dependent permeability may be treated 
without further conceptual difficulty.
The Maxwell equations read as follows:
\begin{eqnarray}
\partial_t B_\alpha &=& -\epsilon_{\alpha\beta\gamma} \partial_\beta E_\gamma\label{induct.eq}\\
\partial_t E_\alpha &=& \epsilon_{\alpha\beta\gamma} c^2\partial_\beta B_\gamma + S_\alpha,\\
S_\alpha &=&  c^2 \epsilon_{\alpha\beta\gamma} \,B_\beta \,\partial_\gamma\ln\mu_0 \nonumber\\
&-&\frac{j_\alpha}{\epsilon_0} - \int_0^t dt'\, \dot\epsilon(\bx, t-t')\, E_\alpha(\bx,t'),\\
&&\partial_\alpha B_\alpha=0,
\end{eqnarray}
where we use Einstein's summation convention, $\epsilon_{\alpha\beta\gamma}$ is the Levi-Civita tensor, $E_\alpha$ is the electric field, $B_\alpha$ the magnetic field, $j_\alpha$ the current, 
$\mu_0(\bx), \epsilon_0(\bx), \epsilon(\bx,t)$, are permittivities that may be inhomogeneous, 
$\dot\epsilon \equiv \partial_t\epsilon$, $\bx$ and $t$ are spatial and temporal coordinates respectively, 
speed of light $c(\bx) = 1/\sqrt{\mu_0\epsilon_0}$, and $S_\alpha(\bx,t)$ is the source term 
into which all inhomogeneities are collected. The temporal convolution term is a
causal model of frequency-dispersive permittivity and is termed the {\it polarization} field. 
In deriving these equations, we have split the electric displacement
field, which is a sum of electric and polarization fields, into two terms and treat 
the latter as a source. 
We do so in order to bring the Maxwell equations into a 
conservative form, amenable to solution by LB methods. 

Various numerical techniques may be utilized to solve these integro-differential equations - one of the 
most popular is to treat the convolution in terms of auxiliary differential equations (ADEs) (e.g., \cite{ADE}).
As in standard control theory, the Laplace transform of $\epsilon(\bx,t)$ may be written 
in terms of its zeros and poles \cite{ADE2} and using these, a system of differential equations,
whose effective continuous-time response is the convolution, may be constructed. 
A similar method may be used in order to address frequency-dispersive magnetic permeabilities.

Although conceptually straightforward, the computational treatment of these 
convolutions is rather laborious, and consequently, we shall leave it for a future separate study.
In what follows, we focus on a simpler case in which dielectric and magnetic permittivities are 
allowed to vary spatially and but are temporally stationary.

A key step to developing an LB formulation is to cast the Maxwell equations in conservative form.
To this end, we express the curl of electric field as the
divergence of an anti-symmetric second-order tensor, 
$\lam_{\alpha\beta}= -\epsilon_{\gamma\alpha\beta} E_\gamma$, such that
$\partial_\beta \lam_{\beta\alpha} = \epsilon_{\alpha\beta\gamma}\partial_\beta 
E_\gamma$.
As a result,  the induction equation~(\ref{induct.eq}) reduces to
$\partial_t B_{\alpha} + \partial_{\beta} \lam_{\beta\alpha} = 0$.
By invoking the Levi-Civita identity,  
$\frac{1}{2}\epsilon_{\gamma\alpha\beta}~\epsilon_{\tau\alpha\beta}, 
= \delta_{\gamma\tau}$, the electric field may be recovered through:
$E_\gamma = -\frac{1}{2}\epsilon_{\gamma\alpha\beta} \lam_{\alpha\beta}$.
In terms of new variables $B_{\alpha}$ and $\Lambda_{\alpha\beta}$,
the Maxwell equations take on the following conservative form:
\begin{eqnarray}
\partial_t B_\alpha + \partial_\beta\lam_{\beta\alpha} &=& 0,\label{lb.bfield}\\
\partial_t \lam_{\alpha\beta} +  c^2 \Omega_{\alpha \beta} &=& -{\epsilon_{\gamma\alpha\beta}}{}S_\gamma\label{lb.efield},
\end{eqnarray}
where $\Omega_{\alpha\beta} 
\equiv \partial_{\alpha} B_{\beta}-\partial_{\beta} B_{\alpha}$ is the 
curl of the magnetic field 
(the wedge curl $\nabla \wedge \bb$ in Clifford algebra terminology).
The wedge curl of the magnetic field may also be expressed as the divergence of a 
triple-rank tensor, so that the Maxwell equations appear in fully conservative form but
we are able to avoid this additional complication. {{Note that if $\partial_\alpha B_\alpha = 0$
at $t=0$, then it will remain so for all time since the partial derivative with respect to $\alpha$ of equation~(\ref{lb.bfield}) 
gives $\partial_t\partial_\alpha B_\alpha= - \partial_\alpha\partial_\beta \lam_{\beta\alpha}$, or, $\partial_t\partial_\alpha B_\alpha = -\epsilon_{\gamma\beta\alpha}\partial_\alpha\partial_\beta E_\gamma \equiv 0$.}}

\section{Lattice BGK formulation}

The next step is to formulate a discrete kinetic equation, whose continuum
limit reproduces the Maxwell equations in the form (\ref{lb.bfield}),~(\ref{lb.efield}) given above. 
To this end, we define the pseudo-vector
distribution function $\bdist_i \equiv g_{i\alpha\beta}$, 
where $(\alpha,\beta)$ are vector indices and $i$ labels the 
discrete identity of the particle moving with velocity $\bvel_i$, to be detailed shortly.
 
The pseudo-vector distribution is postulated to obey the LB equation in BGK form
\begin{eqnarray}
\label{LBE}
\Delta_i \bdist_i &\equiv& \bdist_i (\bx + \bvel_i\Delta t,t+\Delta t)-\bdist_i(\bx,t)\nonumber\\&=&
-\frac{\bdist_i - \bdist_i^{(0)}}{\tau}  \Delta t + {\bf T}_i \Delta t,\label{dist.func.eq}
\end{eqnarray}
where we define the tensor source term as:
${\bf T}_i \equiv
{T}_{i\alpha\beta}= -\frac{w_i}{c^2}\epsilon_{\gamma\alpha\beta} S_\gamma$.
The local equilibrium is chosen so as
to reproduce equations~(\ref{lb.bfield}),~(\ref{lb.efield}) in the continuum, and
it reads as follows:
\begin{eqnarray}
\label{distrib.func}
\dist^{(0)}_{i\alpha\beta} &=& \frac{w_i}{c^2}\left[\lam_{\alpha\beta} + 
c_{i\alpha} B_\beta - c_{i\beta} B_\alpha \right],
\end{eqnarray}
where $c_{i\alpha}$ and $w_i$ are discrete
particle velocities and lattice weights, respectively. 
The structural difference with hydrodynamics is apparent here: inner scalar
products are replaced by outer (wedge) products.

In $D$-dimensional space, these equations are formulated 
in a nearest-neighbor lattice with $2D$ discrete speeds, plus a rest particle. 
For instance, in two spatial dimensions and counterclockwise counting, we 
have $\bvel_1 = (1,0)$, $\bvel_2=(0,1)$, $\bvel_3=(-1,0)$,
$\bvel_4=(0,-1)$ and $\bvel_0=(0,0)$.
The non-moving rest particle `0' ($\bvel_0 = (0,0)$), is instrumental in implementing 
a spatially varying wave propagation speed.
To this end, weights of moving particles are all set to the 
value $w_i(\bx)=(1-w_0(\bx))/2D$. 

According to standard LB theory (e.g., \cite{CHOPA}, \cite{MARCONI}), wave propagation speed (squared) is 
given by the weighted sum of squared particle speeds, 
$c^2=\sum_i w_i c_i^2$.
As a result, for the $6+1$ speeds lattice, we obtain  $c^2(\bx)=(1-w_0(\bx))/3$, which is the 
desired spatial dependence of lattice light speed. 

Note that by choosing different weights for
particles moving along different directions $x$,$y$ and $z$, 
we are able to model anisotropic media.

The first three moments of the equilibrium distribution 
function (Eq.~[\ref{distrib.func}]) are readily computed
\begin{eqnarray}
G_{\alpha\beta}^{(0)} \equiv \sum_i g^{(0)}_{i\alpha\beta}=
\frac{\lam_{\alpha\beta}}{c^2}\label{relate.1},\\
G_{\alpha\beta\gamma}^{(0)} \equiv \sum_i \vel_{i\gamma}g^{(0)}_{i\alpha\beta} 
= \delta_{\gamma\alpha}B_\beta - \delta_{\gamma\beta}B_\alpha\label{relate.2},\\
G_{\alpha\beta\gamma\kappa}^{(0)} \equiv \sum_i \vel_{i\gamma}\vel_{i\kappa}g^{(0)}_{i\alpha\beta}
= \delta_{\gamma\kappa} \lam_{\alpha\beta}.\label{relate.3}
\end{eqnarray}
We also stipulate that the distribution function (Eq.~[\ref{distrib.func}]) 
satisfies \lq \lq mass-momentum" conservation constraints
\begin{equation}
\label{CONSERV}
\sum_i (g_{i\alpha\beta} - g^{(0)}_{i\alpha\beta}) \phi_i \; = \; 0,
\end{equation}
with $\phi_i \equiv (1,c_{i\gamma})$. 

{{The above constraint implies that the
non-equilibrum component of $g_{i\alpha\beta}$ must contribute zero change 
to local mass and momentum, which is a distinctive feature of model 
BGK equations in general}}.

{ Higher-order conservations 
may also be enforced, but these require the use of more
complex lattices, with higher symmetries.
Although certainly within the realm of possibility, it is more labor intensive, especially
in connection with complex geometries. 
}

To analyze the continuum limit, we Taylor expand the left side (streaming operator) of
equation~(\ref{LBE}) to first order in $\Delta t$, to obtain:
$\Delta_i \sim  \Delta t D_i \equiv  \Delta t (\partial_t + \bvel_i \cdot \nabla)$, where 
$D_i$ is the Lagrangian derivative along the $i$-th discrete direction.  
Summing over lattice speeds and invoking constraints~(\ref{CONSERV}), 
\begin{eqnarray}
\partial_t G_{\alpha\beta} + \partial_{\gamma} G_{\alpha\beta\gamma}=-\frac{1}{c^2}\epsilon_{\gamma\alpha\beta}S_\gamma\label{GMOM.1},\\
\partial_t G_{\alpha\beta\gamma} + 
\partial_{\kappa} G_{\alpha\beta\gamma\kappa}=0\label{GMOM.2},
\end{eqnarray}
where, by definition, $G_{\alpha\beta} \equiv \sum_i g_{i\alpha\beta}$,
$G_{\alpha\beta\gamma} \equiv \sum_i g_{i\alpha\beta} c_{i\gamma}$,
$G_{\alpha\beta\gamma\kappa} \equiv \sum_i g_{i\alpha\beta} c_{i\gamma} c_{i\kappa}$.
By construction, $G_{\alpha\beta}=G_{\alpha\beta}^{(0)}$
and $G_{\alpha\beta\gamma}=G_{\alpha\beta\gamma}^{(0)}$, which, along with the
approximation
$G_{\alpha\beta\gamma\kappa} \sim G_{\alpha\beta\gamma\kappa}^{(0)}$, 
and accounting for~(\ref{relate.1}),~(\ref{relate.2}),~(\ref{relate.3}), allows us
to rewrite moment equations~(\ref{GMOM.1}),~(\ref{GMOM.2})
\begin{eqnarray}
\partial_t \lam_{\alpha\beta}+ c^2(\partial_\alpha B_\beta - \partial_\beta  B_\alpha) &=& -\epsilon_{\gamma\alpha\beta} S_\gamma,\label{FINAL.1}\\
\delta_{\kappa\alpha}\partial_t B_\beta - \delta_{\kappa\beta}\partial_t B_\alpha+ \delta_{\kappa\gamma}\partial_\gamma \lam_{\alpha\beta} &=& 0.\label{FINAL.2}
\end{eqnarray}
It may be verified that~(\ref{FINAL.2}) for $\beta = \kappa \neq \alpha$ (the case $\alpha=\beta$ being trivial) returns~(\ref{lb.bfield}).
In summary, moment equations~(\ref{FINAL.1}),~(\ref{FINAL.2}) are shown to reduce to
the Maxwell equations in conservative form. 

As is well known in the case of fluids, a consistent analysis of 
dissipative terms requires the streaming operator to be expanded 
to second order, i.e.,  $D_i = \partial_i + \frac{\Delta t}{2} \partial_i$.
Lengthy algebra shows that the right side of equation~(\ref{FINAL.2}) acquires 
a dissipative term of the form $c^2(\tau-\Delta t/2) \Delta (\partial_{\beta} \lam_{\alpha\beta})$. 
As discussed in \cite{CHOPA} for the case of scalar waves, this is set to
zero by choosing $\tau=\Delta t/2$ ($\tau = 1/2$ in lattice units).

The above formalism readily translates into a simple and actionable algorithm, consisting
of two basic steps: {\it collision} and {\it streaming}. 
In the former, one first prepares the so-called {\it post-collisional} state as follows:
\begin{eqnarray}
\label{COLLI}
g'_{i\alpha\beta}(\bx,t)&=&(1-\frac{\Delta t}{\tau}) g_{i\alpha\beta}(\bx,t) 
+ \frac{\Delta t}{\tau} g^{eq}_{i\alpha\beta}(\bx,t)\nonumber\\
&+& T_{i\alpha\beta}(\bx,t) \; \Delta t,
\end{eqnarray}
where macroscopic variables such as electric and magnetic fields, needed
to compute local equilibrium, are obtained 
from moment equations~(\ref{relate.1}) and~(\ref{relate.2}).
Subsequently, in the streaming step, the post-collisional state is
simply shifted to a neighboring lattice point depending on the direction
and sign of the velocity, namely: 
\begin{equation}
\label{STREAM}
g_{i\alpha\beta}(\bx + \bvel_i\Delta t,t+\Delta t) = g'_{i\alpha\beta}(\bx,t)
\end{equation} 
It may be verified that equations~(\ref{COLLI}) and (\ref{STREAM}) are strictly 
equivalent to the LB equation~(\ref{LBE}).

The resulting numerical procedure is elegant and easy to code, as thoroughly
discussed in numerous introductory texts on this topic (e.g., \cite{INTRO}).

\section{Numerical results:  2D wave propagation}

Since one of the highlights of our scheme is built-in antisymmetry, 
we first inspect numerical errors in maintaining a divergence-free
magnetic field in a homogeneous medium, where $c^2(x,y) = 1$, the 
lattice speed being measured in units of its uniform value.
Particle streaming and collisions are performed for all components of 
the distribution function on a $4+1$-speed two-dimensional
lattice, and we operate at $\tau = 1/2$ (in lattice units $\Delta t=1$).
The implementation is akin to standard LB (e.g., \cite{AIDUN}), the only difference 
being the inclusion of a multi-component distribution function.  

\subsection{Homogeneous media}

\begin{figure}[!ht]
\centering
\includegraphics*[width=\linewidth]{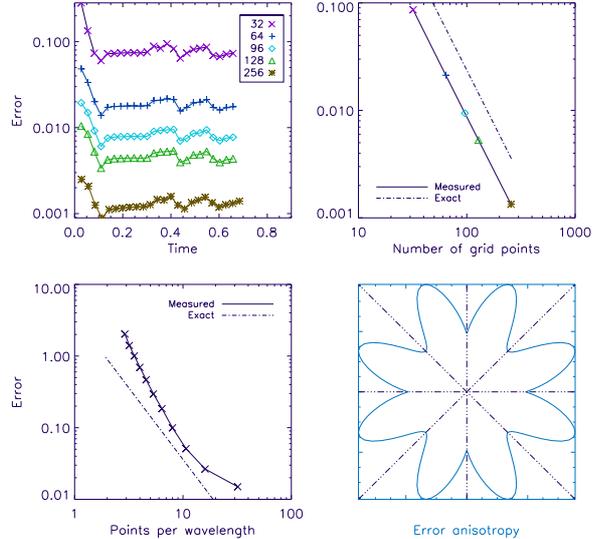}
\caption{Errors in maintaining a divergenceless magnetic field as 
defined by equation~(\ref{error.def}). On the upper
left panel is the error in the $L_2$ norm plotted as a function of time; note that there is an initial 
transient in the magnetic field, due to the applied current, followed by a 
period where the error undergoes some fluctuations. 
On the upper-right panel, the error as a function of grid spacing in 
the $L_2$ norm is plotted, along
with a nominal second-order convergence rate curve. 
For a given calculation ($n_x = 32$ here), the error is
Fourier-decomposed and plotted as a function
of wavenumber or equivalently, points per wavelength on the lower-left panel. 
The nominal line is also plotted, but the error is seen to only approximately 
fall at a second-order rate. 
Lastly, the error is anisotropic, as shown on the lower-right panel. 
Distance from the coordinate center denotes error magnitude; waves propagating 
at 22.5$^\circ$ with respect to the axes are seen to be more inaccurately resolved than in other directions. 
This is a function of the choice of lattice and, in principle, error anisotropy may 
be reduced through the use of higher-order lattices.\label{divb.error}}
\end{figure}

We excite waves by forcing the medium through current density 
$j_z(x,y,t) = \exp\left[-[(x-0.3)^2 + (y-0.4)^2]/({2\times 0.03^2})\right]$, where
$\epsilon = 1$, and $(x,y) \in [0,1)$. All fields are initialized to zero. 
The general wavenumber ($\bk = [k_x,k_y]$) dependent error $\varepsilon$ is described by
\begin{equation}
\varepsilon(\bk) = \frac{\int d\bk'~ {\mathcal F}(\bk,\bk')~ |i\bk'\cdot{\bf B}(\bk')|} { \int d\bk'~ {\mathcal F}(\bk,\bk')~ |{\bf B}(\bk')|  },\label{error.def}
\end{equation}
where we set ${\mathcal F} =1$ to obtain error in the $L_2$ norm, 
${\mathcal F} = \exp\left(- {(|\bk| - |\bk'|)^2}/{2\sigma^2}\right)$ to model the variation of error as a function of absolute 
wavenumber, and thirdly, ${\mathcal F}=\exp\left(- {|\bk - \bk'|^2}/{2\sigma^2}\right)$ to capture 
error dependence on angle of propagation at fixed absolute wavenumber. 
These forms of error are plotted in Figure~\ref{divb.error}. 
{ In general, we expect measures of model accuracy to reflect
a second-order convergence rate. 
For example, the error in enforcing Gauss' law of charge conservation is also likely
to be accurate only to second order (although not confirmed by these tests). 
In the 2D cases considered here, the divergence of the electric field is identically zero
here since ${\bf E} = \{0, 0, E_z(x,y)\} \rightarrow \bnabla\dotp {\bf E} = \partial_z E_z = 0$.}

\subsection{Heterogeneous media}

\begin{figure}[!ht]
\centering
\includegraphics*[width=\linewidth]{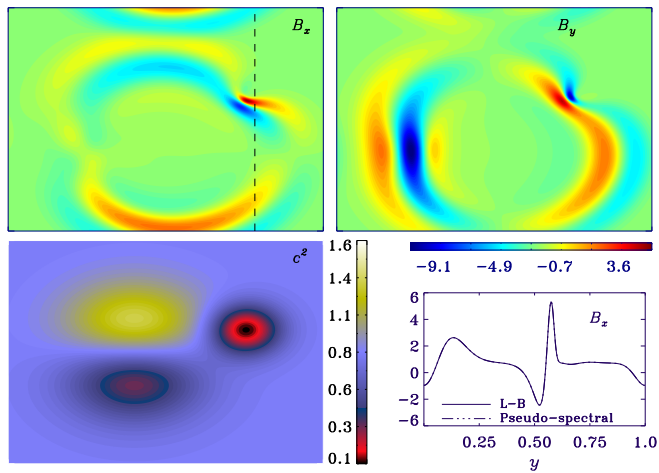}
\caption{Upper panels show snapshots of $B_x$ and $B_y$ at $t=1.272$, with the initial condition described in the text;
the lower left shows the spatial distribution of $c^2$. Axes of the three contour plots are $x$ and $y$.
In both calculations, the grid contained $512\times 512$ points and the domains were horizontally periodic, with $x,y \in [0,1)$.
The horizontal colorbar applies to the upper panels while the vertical bar describes the range in $c^2$.
Regions of low wave-speed (compared with reference $c^2 = 1$; note that $c^2 > 1$ in vacuum, 
although we do not specify a value, since vacuum regions are not included in 
the computational domain) cause waves to refract towards them and vice-versa. 
The lower-right panel compares pseudo-spectral and LB solutions along a cut at constant 
$x$ (location indicated by the dashed line in the $B_x$ plot).
The technique is seen to accurately simulate wave propagation through 
regions where refractive index $n~(\propto c^{-1})$ shows substantial local variations. 
The $L_2$ norm of the difference between LB and pseudo-spectral solutions is 0.6\%.
In order to prevent boundary periodicity from affecting the results, 
a larger computational domain may be chosen and the calculation terminated 
before waves reach the boundaries.}
\label{params}
\end{figure}

Next, we simulate wave propagation in an inhomogeneous medium,
with the initial condition $B_x = (y-0.35)f(x,y)$, 
$B_y = - (x-0.5)f(x,y)$, where $f(x,y) = 10\exp\left\{-[(x-0.5)^2 + (y-0.35)^2]/(2\times 0.04^2)\right\}$.
As may be seen on the lower-left panel of Figure~\ref{params}, the shape, size and 
magnitude of dielectric ``defects" in this calculation bear a qualitative
resemblance with the intermediate matched-impedance zero-index material region (MIZIM) discussed 
in \cite{HET}.   However, different from \cite{HET}, who consider
frequency-dependent permittivity (implying a convolution with electric field), 
we only consider the time-stationary case. 
Also, we do not include inlet and outlet vacuum regions, but implement 
periodic boundary conditions. 
In order to avoid finite-size effects due to recirculating waves, the 
simulation is terminated before waves reach boundaries. 
 
\begin{figure}[!ht]
\centering
\includegraphics*[width=\linewidth]{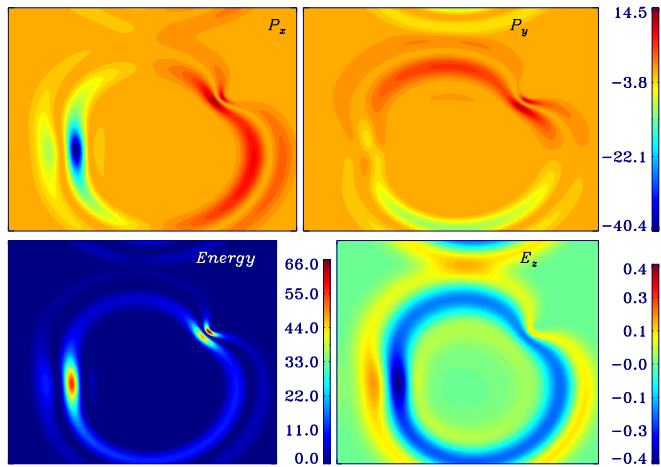}
\caption{Upper panels show the $x-y$ components of Poynting vector, and lower panels display energy and electric field $E_z$.
Axes of all plots are $x$ and $y$, on a periodic grid of $512\times512$ points and $x,y \in [0,1)$.
Since waves spend more time in regions of low wave-speed (shorter wavelengths), 
energy appears to be concentrated in these areas. The top-right scale applies to both upper panels.}
\label{poynting}
\end{figure}

The initial Gaussian wave-packet  splits into up- and downward propagating 
components. 
Waves of finite spatial extent refract into (away from) regions of low (high) $c$,
because the portion of the waveform within the inhomogeneity 
propagates comparatively slower (faster). 
Moreover, since we study linear waves, wave frequency may be regarded 
as invariant, so that wavelength $\lambda \propto c$.
Consequently, waves exhibit locally smaller (larger) wavelengths 
in regions of low (high) $c$. This implies that energy, defined 
as $ {\mathcal E} = \sum_\alpha(B^2_\alpha + E^2_\alpha/c^2)$, tends to concentrate
in regions of low $c$, because waves spend larger fractions of time in these areas.
The $(x,y)$ components of the Poynting energy-flux vector, 
${\mathcal P}_\alpha = \epsilon_{\alpha\beta\gamma} E_\beta B_\gamma$,
and electric field $E_z$ are also shown.
These interpretations are confirmed by visual inspection of properties
of the wavefield and $c^2(\bx)$, as shown in Figures~\ref{params} and~\ref{poynting}. 
Note that the magnitude of $c^2$ in vacuum is necessarily greater than the largest value in this medium. 
The refractive index, $n \propto c^{-1}$, is seen to vary by a factor of four 
($c^2 \sim 0.1- 1.6$), comparable to contrasts used in \cite{HET} to fine-tune
transmission (reflection) coefficients across the MIZIM region.

In order to test the accuracy of the LB solution, we repeat this calculation
using a pseudo-spectral solver, in which spatial derivatives are computed 
spectrally and time stepping is achieved through the application of an 
optimized Runge-Kutta scheme \cite{hu}. 
In Figure~\ref{params}, we also show a comparison between outputs of the two simulations. 
Grids in both solvers contain $512\times512$ points; LB 
and pseudo-spectral solutions are very similar, with a 0.6\% $L_2$ norm of the difference. The LB calculation
at this resolution is approximately 4 times faster than its pseudo-spectral counterpart; however, these codes were
not written with optimization and efficiency in mind and the number may only be interpreted loosely.

\section{Outlook}

In summary, we demonstrate that top-down-prescribed distribution functions of 
\lq \lq particles", not based on any known microscopic kinetic theory of EM,
succeed in reproducing the behavior predicted by the Maxwell equations.
Moreover, our result provides a further example of lattice kinetic theory
as an efficient tool to simulate continuum-physics phenomena via a particle-like formalism,
well suited to handling complex geometries and showing excellent scalability on
modern parallel computers \cite{SIM}.
The present method may be extended in many directions, such as
invisibility cloaks, i.e., systems wherein the use of meta-materials allows for selected regions of space 
to become inaccessible to light,  analogous to metric holes in space-time (e.g., Fig. 3 in \cite{Wegener}). 
Another possibility is to extend it to study plasma phenomena in confined geometries. 

We note potential limitations of this technique and describe directions for future work.
Like most LB methods, the present scheme is formulated in a space-time uniform lattice, so that
space and time resolution may not be changed independently unless locally-adaptive formulations of the
method are put in place. Such adaptive LB formulations do indeed exist \cite{ARLB}, although 
they are usually significantly more laborious than the native version on uniform lattices.   
A second point regards numerical stability in the presence of sharp interfaces, resolved by just a few 
lattice points. At such sharp interfaces, higher order terms may no 
longer be negligible, thereby introducing unwanted dissipative effects. 
Efficient implementations with non-local permittivities, as mentioned 
earlier on in this paper, are likely to require a careful analysis of the auxiliary equations
to be coupled with the native LB equation for waves.
This is similar to the way auxiliary equations are coupled
to the LB equation for the modeling of turbulent fluid flows \cite{SCI}. 

{ The present LB scheme may be regarded as a special type of finite-difference method, in which streaming
is exact and local conservations are built-in and accurate to machine
         round-off. Although the technique is just second-order in 
         space-time, the above properties make the error prefactors 
         sufficiently small so as to render its performance competitive with
         higher-order methods, including spectral ones.  }
         
{ From a mathematical stand point, the inclusion of source terms, reflecting external sources 
and/or inhomogeneities, is straightforward, once the expression of these sources in the 
continuum is known. However, the strengths of such terms may place stringent constraints on the 
stability of the scheme, aspects that remain to be investigated.

Although we have only considered periodic boundaries, we note that
there exists literature on the implementation of other boundary conditions, such
as reflecting, absorbing etc. (e.g., \cite{AIDUN} and references therein)
As a result, the formulation of different types of boundary conditions
for the wave-LB scheme might follow from previous developments although
such a possibility remains to be studied in full detail.
}

Despite their importance, the above developments that do not challenge
the basic merits of the scheme discussed in this paper.
As a result, we believe that the LB scheme presented in this work should offer a fast 
and flexible computational tool to assist, complement and possibly even anticipate experimental research on 
invisibility cloaks and related phenomena in modern optics and photonics research \cite{PENDRY1,PENDRY2}.

\acknowledgements
Computations were performed on the Pleiades cluster at NASA Ames. 
Valuables discussions with J. Wettlaufer (Yale Univ.) and X. Shan (Exa Corp) 
are kindly acknowledged. S.S. thanks P. Dellar (Oxford Univ.) for sharing his insights. 
We acknowledge support from NSF Grant OCI-1005594. 

\bibliographystyle{epl2}

\end{document}